# Industrial Limitations on Academic Freedom in Computer Science

Reuben Kirkham[1[0000-0002-1902-549X]]

[1] Monash University, Victoria, Australia
reuben.kirkham@monash.edu

**Abstract.** The field of computer science is perhaps uniquely connected with industry. For example, our main publication outlets (i.e. conferences) are regularly sponsored by large technology companies, and much of our research funding is either directly or indirectly provided by industry. In turn, this places potential limitations on academic freedom, which is a profound ethical concern, yet curiously is not directly addressed within existing ethical codes. A field that limits academic freedom presents the risk that the results of the work conducted within it cannot always be relied upon. In the context of a field that is perhaps unique in both its connection to industry and impact on society, special measures are needed to address this problem. This paper discusses the range of protections that could be provided.

## 1 Academic Freedom: What is it and why does it matter?

Academic Freedom (otherwise known as Intellectual Freedom) is a civic right of considerable importance. It is about ensuring that individual academics are in the position to question received wisdom and speak truth to power. As such, academic freedom is intimately connected to the role of a University in the 'search for truth' (Hudson & Williams, 2016) and is "*at the very core of the mission of the university"* (Altbach, 2001). This means that a culture where academic freedom is fully supported is necessary to ensure that the knowledge produced by an academic community can be fully relied upon. More than ever (perhaps especially given the COVID-19 pandemic), this is fundamentally important to wider society: academic freedom is a key part of ensuring that the general public can trust and rely upon research findings.

It is difficult to understate how fundamental academic freedom *should* be in academia. It is intended to play "*an important ethical role not just in the lives of the few people it protects, but in the life of the community more generally*" to ensure that academics do "*not [feel compelled] to profess what one believes to be false"* and amounts to *"a duty to speak out for what one believes to be true*" (Dworkin, 1996). Moreover, it has a fundamental societal importance beyond the pursuit of truth itself, with academic freedom also playing a "*valuable role … in supporting democratic government*" (Evans & Stone, 2021).

In our field, existing ethical codes expect us to take an active role in the community and advance 'social good'. For example, the ACM Code of Ethics expects computing professionals (including academics) to "*give comprehensive and thorough evaluations of computer systems and their impacts, including analysis of possible risks*." (2.5), "*foster public awareness and understanding of computing, related technologies, and their consequences*" (2.7) and "*ensure that the public good is the central concern during all professional computing work*" (3.1). Yet, this ethical framework does not directly address academic freedom, and thus does not deliver a culture where these ethical principles will likely be put into practice. Nor has there been a publicized case where the ACM or the IEEE have taken steps to enforce academic freedom, which suggests at the least, a lack of prominence in their organizational thinking in respect of this concern.

The concern of academic freedom is perhaps of particularly strong importance in computer science. It is difficult to think of a research field that has a greater permeation into our day-to-day lives, especially in light of the global pandemic. For instance, computational modelling has been used as a basis for locking down populations, whilst our field has also been essential for enabling people to remain connected whilst socially distancing in the physical world. This means that the work conducted in our field has life-altering consequences for entire populations, making our work carries a particular social importance.

If we as a field are to be worthy of societies trust, then it is of considerable importance to ensure there are high standards of academic freedom. In practice, academic freedom tends to suffer from improper limitations. This means that other people are able to restrict or chill speech, or the research that is conducted. The undermining of academic freedom can come from a range of sources, such as other academics, an academics own institution, public funding bodies, or industry (Evans & Stone, 2021; Hudson & Williams, 2016). This paper focusses on the particular arrangements between industry and academia within the field of computer science and explains how they undermine academic freedom. It also proposes how existing practices should change so the public can have more confidence in the research conducted within our field.

## 2 How might academic freedom be uniquely limited in computer science?

Unfortunately, academic freedom is particularly restricted within computer science, relative to most other fields. One limitation arises from our fields extensive connection to industry, where unlike other fields, employs a large number of the best scientists in quazi-academic roles (Ebell et al., 2021). Just to give an example of the scale of this, Microsoft Research has been described as the "*largest computer science department*" in the world[1], whilst being listed as a partner on over £340 million of EPSRC funding.[2]

[1] https://www.computerworld.com/article/2547042/microsoft-research--at-15--looks-ahead-to-more-innovations.html

[2] See https://perma.cc/L3TP-CQQB

The issue is there is a considerable interconnection between large technology companies and the academic community. For a computer science academic, this means:

1. Industry influences who is awarded research grant funding. This is either by direct provision (e.g. Google or Microsoft grants), or by the increasing expectation of industry 'partners' on research grant applications to public funders (e.g. UKRI, or the ARC).
2. Many other career opportunities, including future employment, are also controlled by industry. This is particularly so for PhD students, many of whom undertake industry-based internships (and are often provided through connections to PhD supervisors). PhD students also usually have a subsequent career in industry after they graduate (instead of an academic career).
3. Academic careers are intrinsically driven by publications – many of these publications are supported by industry, be it by way of collaborations, or the provision of access to datasets or other materials. These resources are regularly controlled by large companies (e.g. Google, Facebook, Microsoft), giving them a direct influence over the publication capacity of many academics.
4. Academic conferences, which have a co-equal status to journals (uniquely in Computer Science) are not only sponsored by large companies, but their employees are routinely involved in the peer-review process, including by acting in an editorial role (e.g. as program committee members or chairs, or as journal editors).

It might be said that industry has a Jekyll and Hyde type relationship with the academic community in computer science. In its benevolent form, this relationship offers a wide range of expertise and opportunities – these are perhaps especially important for PhD students, many of whom will graduate to a well-paid job in industry. Yet, the types of concerns raised above are likely to have a considerable chilling effect on the free speech of computer science researchers and by implication, academic freedom. This is because many of the (particularly powerful) technology companies have the ability to heavily influence academic careers. This is also compounded by the increasing practice of university's to measure publication performance and research grant income, and even to set targets in this regard for academic staff members. At the same time, there is already a lack of a debate culture within certain fields (e.g. see the account in (Button et al., 2015) in respect of HCI or (Kirkham, 2021) for an example of attempts to discourage certain types of research). This means that an important public function is likely to be being chilled on the ground: whilst this is certainly not only by industry (other academics are also often part of the problem), it is one important part of the problem to be addressed.

## 3 What might be done about this?

The overriding issue is ensuring that academic decision making is not compromised by powerful industrial interests, or the 'chilling effect' arising from the power that industry can hold over individual academic careers. The way to address this is to ensure that decisions about individuals' academic careers cannot be influenced by industry. For wider debate, I would propose a number of steps that might be undertaken:

### 3.1 Not measuring industrial funding as a performance measure.

Unfortunately, it is common for grant income to be used as a measure of 'research performance' (and thus to determine whether an academic will continue to be employed), as well as also serving as a means for advancing academic careers in and of itself (e.g. by enabling more research to be done, and thus increasing the volume of publications). This is a major ethical issue in and of itself, even if industry were not involved. Existing competitions for research grant income between researchers have been long known to be a highly flawed measure of research performance (see e.g. (Gillett, 1991)), as well as amounting to a considerable waste of public resources (Domingos, 2022). Furthermore, the existing practice in most grant allocation processes – especially the failure to adopt anonymous blind review – is well known to promote discrimination against certain minority researchers in respect of funding decisions (Witteman et al., 2019). Of course, the existence of discrimination also suggests an element of capriciousness, with the decision makers being vulnerable to making decisions on irrelevant factors, rather than purely the matters they ought to be considering: indeed, it is well recognized that many funding process eshave a "huge amount of randomness" (see e.g. (Grove, 2021)). Furthermore, the idea of a research grant system is itself inconsistent with academic freedom: the process involves 'senior' academics deciding what research other 'junior' researchers will be funded to do and thus amounts to an effective form of censorship of other researchers.

The provision of funds by industry further compounds this problem, making a bad situation worse. The effect of these financial provisions is threefold:

- Individual companies, or individuals within them, have the ability to advance the careers of specific researchers, both by providing funding directly and the prestige and/or status that comes with this funding (e.g. it is currently seen as prestigious to get funding from Google, Microsoft, Facebook etc.). This means they have the power to advantage the careers of researchers who support their commercial interests and agenda over those who do not.
- Individual companies can shape the research agenda of a field, by funding academics to work on specific research directions. This potentially distorts the research that is conducted. This is perhaps especially problematic if there is match funding provided by an academic department (e.g. a small Facebook grant ends up being topped up into being a full PhD scholarship).
- The amount of overall grant funding increases, thus increasing the grant targets imposed on academics overall, and penalizing those who do not wish (or are less able) to obtain industry funding.

In some respects, this is something that can be relatively easily resolved. First, academic institutions should not count industry supported research funding (including from other sources where industry 'partners' were involved in supporting the funding application) as an indication of performance, given the potentially corrosive effects on academic freedom. Second, there should be an effort to eliminate ‘Matthew’ effects, so that if something is funded by industry, then other resources (e.g. discretionary funds) should be diverted into areas that industry is less likely to fund (for example areas that are of social or public importance). This would be different from the present situation, where a ‘profit’ normally gets sent to central funds (e.g. to fund buildings or administration) or alternatively to the research funds of the researcher who obtained the industry funding to begin with (thus rewarding them further), rather than being spent on alternative research. Third, there should be a prohibition on ‘match’ funding: instead, industry funded research should in effect be heavily ‘taxed’ (e.g. by a heavy ‘overhead’ charge) that gets redirected into non-industry related research by other academics.

Industry can take some positive steps too. I would argue that industrial organizations who act ethically should not be giving funding to *named* individual researchers, but instead funding organizations, thus dampening the individual prestige effect. Furthermore, if industry wishes to sponsor research, then a better position would be for this to be done following a *double-blind review* process where the reviewers are independent of the relevant corporate interests. There is nothing wrong with industry funding an area of research, provided that this funding is done in a manner that is conservative of academic freedom. The problem is that industry funding is presently being used in an inappropriate manner, and academic institutions have not put in place measures to deal with this problem.

A more radical perspective could be to take a more restrictive approach. Many academic institutions would not, for example, accept research funding from a ‘tobacco’ company (Thomson & Signal, 2005): this is due to the harm such research can impose on public health (Turcotte, 2003). Might there not be room for a similar approach towards certain technology companies that seek to fund research in University’s, whilst not fully respecting academic freedom? At the least, bodies like the IEEE and the ACM could make it a requirement that they will only accept papers funded by an appropriate source, i.e. where the underlying funding has been provided in an ethical manner which fully protects academic freedom.

### 3.2 Removing the role of industry in publicly funded grant competitions.

There is a related issue of concern. It is common for public bodies that fund research to favour applications that have already had industrial support (e.g. someone has had awards, collaborations or funding from Microsoft, Google or Amazon or other companies), or by the provision of future in-kind industry support with respect to a grant application. The purpose is to ‘translate’ academic research into tangible outcomes and benefit the economy. The difficulty is that the present approach of doing so risks undermining academic freedom, by providing a mechanism by which industry can favor certain academics, or their institutions. Beyond the direct industry influence, such a process is also unfair more generally, in that it fails to follow a *double-blind review* process, something which is well known to discriminate against minority groups, as well as itself risking academic freedom. This risk is because an anonymous reviewer – which might themselves have some connection to industry, or dislike another exercise of academic freedom - can reject a proposal without accountability based on the identity or activities of the applicant, rather than the merits of their research.

Addressing this issue requires a focus on the requirement that is being imposed on a researcher, namely that the researcher themselves has a pre-existing *relationship* with a given sponsor, and this relationship helps their chances (or is a hard requirement to succeed) in these competitions. It is the *a-priori* quality of the relationship that is the problem: if there were a structured means for involving industry *after* the grant has been awarded, then the difficulty would likely disappear. There are various *post-hoc models* that can bring this about, whilst conserving academic freedom. For example:

- Having a panel of industrial organisations, who are allocated the most appropriate projects on a *post-hoc* basis, *after* an award/funding decision has been made.
- Placing more of an emphasis on start-ups and small-businesses (instead of large companies), which considerably reduces the ethical risk (these organizations could also be anonymized in the peer-review process).
- Supporting academics to start their own companies and startups, thus mostly sidestepping existing technology companies.

.
Adopting any of these models would be in the public interest. By doing it on a *post-hoc* basis, it is also arguable that collaborations would be better formed: rather than being based on who is already ‘connected’, the most appropriate industry organisation can later be assigned as the project partner. Industry and academic time would be saved, because only the (small proportion) of grant proposals which are eventually funded would need partnerships to be arranged. Furthermore, if configured appropriately, smaller and medium sized businesses would benefit, due to the reduction of unfair competition (e.g. researcher applicants don’t need a recognized ‘name’, but can focus on partnering with the most appropriate organizations when they have the funding down the track). The wider benefits, and the likelihood of this better serving industry (especially by mitigating the present bias in favour of large international tech companies), enhances the ethical case for reform in this regard. To put it another way, with carefully thought-out processes, the involvement of industry in supporting research can co-exist with academic freedom: the problem is the absence of appropriate policies.

### 3.3 Reducing the unfair influence of industry resources

For certain favoured academics, industry provides them with a range of resources, which helps advance their research. This does not only include funding, but also access to expertise, existing trade secrets, software tools, and data. In many cases, especially in respect of large technology companies, it is possible that certain studies can only be done with this level of access or assistance – for instance, a study might need the ability to control the platform in question (e.g. by tweaking the data presented to a proportion of end users), or require access to the enormous amount of data available to large corporations in respect of their own software platforms. To put it another way, many experiments on Facebook (and other platforms) require the consent and co-operation of Facebook itself.

One issue is that the provision of these resources means that some academics will have an advantage over those who are less favored by these companies. There is a real risk that the power that these companies have can influence both the research, and the researcher. This is because of the possibility of this support being withdrawn, which amounts to a chilling effect. In turn, this means it is difficult to have confidence in the research, or indeed researchers who are associated with such matters. Yet at the same time, industry is substantially contributing to research by:

- Providing additional research resources that would otherwise not exist, thus increasing the volume and extent of the research that is done.
- Enabling research that can only be uniquely supported by industry.
- Providing opportunities for researcher development, such as internships for PhD students.

The issue is the existing lack of a governance framework that protects against the underlying chilling effect that arises from the influence of industry. The starting point should be that academics are not rewarded in their careers due to having industry connections (after all, networking ability and ones 'connections' in general has nothing to do with intrinsic academic merit, so rewarding this type of thing is arguably inappropriate to begin with), and the incentive models should be properly configured to protect against this. For instance, although papers supported by large companies should be reviewed and published, I would argue that they should not count as full publications for the purposes of academic promotions or other performance measures. Furthermore, active steps should be taken to remove the perceived 'prestige' of engaging with industry – it should be a matter of free-choice for individual academic staff if they engage (and if they do so, to what extent). Treating industry engagement as particularly prestigious is an inappropriate derogation from the principle of academic freedom, as it undermines a researcher's individual freedom to *not* engage with it (or to engage only on limited terms).

### 3.4 Providing effective information access rights

There is a more fundamental concern that follows on from the foregoing one. This issue concerns a particular type of resource: information. I argue that there should be generally free access to a suitably-qualified academic to the operation of systems by large-scale technology companies. Instead of being a privilege for the 'chosen few' selected by industry (i.e. the status quo), whether a given academic has access should not be based on the patronage of a company (or the networks of that academic), but instead by way of provision made through a fair, merit-based and independent process. This process would mean that the legitimate interests of the company are protected, such as trade secrets and data security, whilst bona-fide academics have the right to conduct appropriate investigations, even if they are investigations the technology companies might find inconvenient.

It is notable that tech companies are particularly active when claiming to be 'good' corporations, with Google's original motto famously being 'do no evil' (Crofts & van Rijswijk, 2020). In a modern democratic society, transparency is a recognized virtue. It is also an ethical standard that many in computing adhere to, perhaps exemplified by the 'open source' movement. At the same time, there are legislative expectations – transparency is an expectation associated with the GDPR (Wachter, 2018). Furthermore, when applied to public institutions – freedom of information is said to be an important "*democratic right*" (Walby & Luscombe, 2019) and given the role of some large technology companies (with some even having a market capitalization that goes beyond most countries annual GDP), there seems to be little reason why such information access rights should not apply to their operations too. One might go as far as to argue that the enhanced scrutiny that fair information access provides should be welcomed by those tech companies who like to impliedly assert that they are paragons of ethics. After all, a tech company acting sincerely should not be able to (or wish to) constrain research access to only 'friendly' researchers.

The ideal solution would be the expansion of freedom of information law to apply to these organizations as if they are public sector organisations: the problem is this is something that requires legislation (as well as an effective enforcement structure that FOI law tends to lack (Worthy, 2017)). However, there is a step that can be taken in respect of the underlying academic freedom issue: if there was not fair access to the underlying resource, then academic conferences and journals should refuse the submission. There are two reasons why this policy would be appropriate. The first is that the academic conference or journal cannot be sure of the reliability of the results, due to the lack of replicability, independence from the prevailing corporate interests, and the lack of any realistic ability to verify the data (for instance, if there was 'cherry picking' it would be very hard to investigate this). There is an obvious ethical risk in imbuing such work with the imprimatur of well-recognized publication venues, especially where industry has a particular interest in certain results and the endorsement by the conference or journal thus supports a particular commercial goal, rather than the public interest. The second is that the industry access is unfair to account for in individual careers: in other words, the industry involvement prevents a fair and meritocratic competition between academics, and also undermines academic freedom (for the reasons given in respect of 3.3). In effect, sufficient openness should become the price of entry into the academic community.

### 3.5 Adopting special considerations for papers that are designed to uncover wrongdoing.

Whether by accident or design, some research investigations end up uncovering improper practices, be it by industry, or other actors, such as parts of the state (e.g. as in the Post Office case in the UK (Wallis, 2021)). Alternatively, this work might uncover inadvertently problematic practices, but those which when identified, would have serious consequences for individual tech companies (for example a security bug in a computer chip). It follows that particular publications may have considerable commercial implications for some organizations, and thus advantage or penalize particular industry players (depending on what the research has discovered). Yet there is an absence of a specialized procedure for reviewing such papers, even though they should be given particular attention for the following reasons:

- The risk of conflict of interest, which could either be in favour of or against acceptance of a particular work, depending on whether the works specific outcomes favour a particular industrial organizations interests. A conflict of interest risks both reputational damage (and thus undermines the ethical imperative of public confidence in science), as well as the substantive fairness of the peer-review process (i.e. a risk of a unfair outcome). The involvement of industry in the determination of paper acceptance further amplifies this risk.
- The increased negative consequences of a false-positive acceptance, given the potential harm an error can cause to wider society. In short, special care is needed, to ensure the accuracy of this work. Peer-review is not always effective in accurately or rigorously identifying errors, so there is an enhanced concern with such works.
- The increased negative consequences of a rejection, which in most cases amounts to a delay (as the authors resubmit the work). In some cases, delayed publication may as well as be denied publication – the delay may undermine the underlying social impact that the work would have otherwise have had. This issue is a well-known consideration in the context of Freedom of Information (where delayed information is often tantamount to denied information), and has equal force in this particular context.
- The fact that in some cases, the authors may in effect be whistleblowers, yet the publication process operates on the basis that the authors are not (for example) anonymous. There is one journal designed to deal with this issue – namely the Journal of Controversial Ideas, which allows anonymous paper authorship (McMahan et al., 2021) – but this is not necessarily configured to deal with matters of computer science, nor can it deal with most such cases that arise in our field.

Perhaps surprisingly, there is no such process at present in our field. Yet it would not be particularly difficult to establish one – it merely requires having a separate track with appropriate provisions and for the resources to conduct this enhanced (and more rapid) peer-reviewing to be provided. The only challenge would be funding – there is a likely need to pay reviewers so that decisions can be made with alacrity and to the required standard, but there is also a likelihood that reviewers may be more willing to do this type of reviewing in any event. However, this simply means that some resources within bodies such as the ACM or the IEEE need to be diverted to support this scheme, compared to other discretionary activities (e.g. by reducing the expenditure on the more lavish aspects of existing conferences entertainment packages).

### 3.6 Requiring industrial related research to undergo special ethical review

There has been a history of problematic studies being conducted by industry, perhaps most infamously the ‘Facebook contagion’ study, as analysed in (Grimmelmann, 2015). This history demonstrates that there are significant ethical risks in respect of experiments conducted by large technology companies to human subjects. However, even if these organisations were to be subject to an independent IRB or other ethical review in respect of their own research activity (which they mostly are not), wider ethical risks remain that go beyond the direct subjects of experiments. Indeed, these wider ethical risks – i.e. social impacts – are the main concern. This means that a special ethical review process is arguably required that addresses these wider risks. In particular, such a process should consider matters such as:

- Does the industrial relationship risk the independence of the research?
- Was there a fair process for obtaining access to data and industrial resources?
- Are the resources that have been provided skewed in some manner, thus allowing the sponsoring company to shape the findings that arise, or lead to ‘uncomfortable’ questions being avoided within the research?
- Is there a risk of a corporate advantage being gained that has a wider ethical risk (e.g. by preventing fair competition, undermining individual consumer interests, or deflecting regulatory oversight)?
- Is there a risk to academic freedom, be it directly or indirectly?

These types of questions arguably should also be considered by a body that is genuinely independent of both the University and Industry. This is another way that such a process may need special constitution – existing IRB’s and ethical bodies are often not independent of the University, and thus may be at risk of being improperly pressured. Ideally, such matters should be considered by an independent administrative tribunal, much in the way that ‘freedom of information’ cases are considered. However, there would be a need to avoid the particular risks that arise with some administrative tribunals (e.g. the risk of bias in favour of political interests (Ellis, 2013)), as well as a lack of the relevant expertise and understanding of information technology (e.g. as with the UK’s information rights system (Kirkham, 2018)).

### 3.7 Requiring the support of academic freedom as a pre-requisite for participation

This final proposal is a more holistic one. Industrial organizations benefit from and participate within our academic community in a manner that does not tend to happen in other disciplines. Our system currently provides a range of soft support to industry, including access to our PhD students and graduates, the review of work conducted by industry-based professionals, and access to expertise within academia.

If they engage with and benefit from our community, then I would argue that we should expect in return that they respect the ground rules. As a minimum, this means a proper respect for the principle of academic freedom. Unfortunately, the academic freedom of industry-based researchers has not always been respected. A recent prominent example is the case of Timnit Gebru and others, which putting aside other events, involved Google requiring the withdrawal of a paper submission ostensibly for perceived quality grounds (Ebell et al., 2021), and therefore undermining a key function of the academic peer-review process, whose role is to decide the merits of individual work. If Google – by arrogating the peer review function to itself – acts as a censor of submissions, then it is arguably unclear how it would be appropriate to allow submissions from researchers based at that organization, or to allow its senior researchers sit on program committees. Perhaps predictably, one effect of the Gebru case was to damage Google's relationship with the academic community, with some conferences refusing to accept funding from them (Johnson, 2021).

It is arguable that an industry body that does not respect academic freedom should not be allowed to make paper submissions, have its researchers sit on program committees, or be otherwise connected to the academic community. Furthermore, given the ACM and IEEE code's focus on individual conduct, one presumably expects that an academic in a University should not be supporting industry organizations who are insufficiently respectful of academic freedom, as it is difficult to see how doing so would amount to "*highest standards of integrity, responsible behavior, and ethical conduct in professional activities*" (per the IEEE code), or be "*encourag[ing the] acceptance ... of social responsibilities by members of the organization"* (per the ACM code). There is a caveat on this, in that the IEEE or ACM has not issued explicit guidance about how academics should (or more accurately should not) engage with industrial organizations who do not respect academic freedom, but as soon as this is spelled out, one would expect that computing professionals in academia would be careful to limit their engagement with them.

## 4 Conclusion

As a field, computer science is perhaps uniquely enmeshed with industry, which poses certain societal risks. This paper has identified a range of vectors by which industry is likely having a negative impact on academic freedom. This means that the part of academia that addresses matters of computer science is not always fulfilling one of its important public functions. This issue needs to be addressed for the good of wider society.

There is one more concluding remark that I should make. This paper deals with only one aspect of concern in respect of academic freedom: there is a wide range of other threats to academic freedom. For example, higher education institutions and academic researchers within them are also problematic in certain ways: this article of course, does not primarily focus on resolving these issues. In particular, there has been a lack of respect of academic freedom by some higher education institutions, with the Peter Ridd and Gerd Schröder-Turk cases in Australia serving as prominent recent examples (Evans & Stone, 2021). There have also been occasions where academics have engaged in campaigns or petitions aimed at undermining the freedom of other academics, with examples including the case of the computational geophysicist Dorian Abbot, who was subject to a campaign for politely expressing his views in support of 'Merit, Fairness and Equality'. Within computing, even the existing peer review process is concerning even without the industrial issue – for example, there have been attempts to accept or reject papers based on perceived ideological viewpoints (e.g. as discussed in Kirkham, 2021), and to engage in what is asserted to be 'citational justice' (see e.g (Collective et al., 2021) for an concerning example of this), where the identity of the author is relevant, rather than the substantive contents of their research. The issue of academic freedom is a matter of wider importance, and it is important not to lose sight of the fact that industry is just one of the existing problems, albeit still a serious one that needs to be addressed as part of a wider debate. This paper aims to help start that debate.